\documentclass[twocolumn,showpacs,preprintnumbers,
                amsmath,amssymb,eqsecnum,aps]{revtex4}
\usepackage{epsfig}
\usepackage{graphicx}% Include figure files
\usepackage{dcolumn}% Align table columns on decimal point
\usepackage{bm}% bold math

\begin{document}

%\preprint{APS/123-QED}

%%%%%%%%%%%%%%%%%%%%%%%%%%%%%%%%%%%%%%%%%%%%%%%%%%%%%%%%%%%%%%%%%%%%%
%%%%%%%%%%%%%%%%%%%%%         Title       %%%%%%%%%%%%%%%%%%%%%%%%%%%
%%%%%%%%%%%%%%%%%%%%%%%%%%%%%%%%%%%%%%%%%%%%%%%%%%%%%%%%%%%%%%%%%%%%%

      \title{
 %\begin{flushright} {\small IFT-P.  \,\, gr-qc/0301141} \end{flushright}
             Weak decay of swirling protons and other processes
            }

%%%%%%%%%%%%%%%%%%%%%%%%%%%%%%%%%%%%%%%%%%%%%%%%%%%%%%%%%%%%%%%%%%%%%       
%%%%%%%%%%%%%%%%%%%%     Authors & Addresses  %%%%%%%%%%%%%%%%%%%%%%%
%%%%%%%%%%%%%%%%%%%%%%%%%%%%%%%%%%%%%%%%%%%%%%%%%%%%%%%%%%%%%%%%%%%%%

\author{George E.\ A.\ Matsas}
\email{matsas@ift.unesp.br}
\affiliation{Instituto de F\'\i sica Te\'orica, Universidade Estadual Paulista,
                 R. Pamplona 145, 01405-900, S\~ao Paulo, SP, Brazil}

\author{Daniel A.\ T.\ Vanzella}
\email{vanzella@uwm.edu}
\affiliation{Physics Department, University of Wisconsin-Milwaukee,
        1900 E Kenwood Blvd., Milwaukee, WI 53201}

\date{\today}% It is always \today, today,

%%%%%%%%%%%%%%%%%%%%%%%%%%%%%%%%%%%%%%%%%%%%%%%%%%%%%%%%%%%%%%%%%%%%%%%
%%%%%%%%%%%%%%%%%%%%%           Abstract         %%%%%%%%%%%%%%%%%%%%%%
%%%%%%%%%%%%%%%%%%%%%%%%%%%%%%%%%%%%%%%%%%%%%%%%%%%%%%%%%%%%%%%%%%%%%%%

\begin{abstract}

We investigate the weak-interaction emission of spin-$1/2$ fermions 
from decaying (and non-decaying) particles endowed with uniform circular 
motion. The decay of swirling protons and the neutrino-antineutrino emission 
from circularly moving electrons are analyzed in some detail. The 
relevance of our results to astrophysics is commented.

\end{abstract}

\maketitle

%%%%%%%%%%%%%%%%%%%%%%%%%%%%%%%%%%%%%%%%%%%%%%%%%%%%%%%%%%%
%!!!!!!!!!!!!!!!!!!!!  Introducao  !!!!!!!!!!!!!!!!!!!!!!!%
%%%%%%%%%%%%%%%%%%%%%%%%%%%%%%%%%%%%%%%%%%%%%%%%%%%%%%%%%%%

\section{\label{intro} Introduction}
Recently~\cite{MV} the decay of uniformly accelerated protons as described 
from the point of view of inertial and coaccelerated observers was 
used as a paradigmatic example 
of the necessity of the Fulling-Davies-Unruh effect~\cite{FDU} to the 
consistency of  Quantum Field Theory. 
At the same time, it was shown that in addition to its conceptual importance,
the decay of accelerated protons could be of ``practical'' 
relevance to astrophysics. It was estimated that about 1\% of
a bunch of  protons with energy $10^{14}$~eV would decay 
(through weak interaction)
if they were
under the influence of a magnetic field  of $10^{14}$~Gauss of a  
pulsar~\cite{VM3}. The proton decay can be
understood in this case as being induced by the centripetal force 
acting on the proton
as it swirls around the magnetic field
lines. The estimative above was obtained, however, by using the decay rate of
{\em uniformly accelerated} protons 
rather than circularly moving ones. It was argued that this procedure
should lead to good approximate results as far 
as the proton proper acceleration satisfies the constraint 
$a \gg \Delta M, 1/R$, where $\Delta M \equiv M_n - M_p$ 
is the neutron-proton mass difference and $R$ is the local 
curvature radius of  the proton trajectory. 

Thus, as a step further, it would be desirable to refine our 
previous estimative by considering  protons in 
circular motion indeed. For this purpose, here we apply the formalism 
developed in Ref.~\cite{VM3} (designed to study the weak-interaction 
emission of spin-$1/2$ fermions from classical 
and semiclassical currents) to the case of circularly moving 
particles with constant velocity  (hereafter denominated 
{\em uniformly swirling particles}). 
We focus on the decay of uniformly swirling protons 
and on the neutrino-antineutrino emission from uniformly swirling 
electrons which is also relevant in some astrophysical situations 
as, e.g., in the cooling of neutron stars 
and in connection with high-energy neutrinos emitted from the
magnetosphere of pulsars~\cite{BK}-\cite{LP}. 
We emphasize that a broad class of 
interesting processes involving accelerated particles possessing
well defined world lines can be analyzed in this fashion.

The paper is  organized as follows: In Sec.~\ref{formalism} 
we present the main results which will be useful for our 
purposes. In Sec.~\ref{current} we explicitly evaluate the fermion 
emission rate and radiated power from uniformly swirling particles,
where we assume Minkowski spacetime with 
metric components $\eta_{\mu \nu} = {\rm diag} (+1, -1, -1, -1)$
associated with the usual inertial coordinates $(t, {\bf x})$. 
In Sec.~\ref{proton} we analyze in 
detail the decay of uniformly swirling protons and comment 
on its potential importance to astrophysics. Sec.~\ref{neutrino}  
is dedicated to analyze the neutrino-antineutrino emission from 
uniformly swirling electrons. We dedicate Sec.~\ref{discussion} 
to our final discussions. We  adopt natural units 
$c = \hbar = 1$ throughout this paper unless stated otherwise.

%%%%%%%%%%%%%%%%%%%%%%%%%%%%%%%%%%%%%%%%%%%%%%%%%%%%%%%%%%%%%%%%%%%%%%%%
%%%%%%%%%%%%%%%%%%%%%%%%%%%  The Formalism  %%%%%%%%%%%%%%%%%%%%%%%%%%%%
%%%%%%%%%%%%%%%%%%%%%%%%%%%%%%%%%%%%%%%%%%%%%%%%%%%%%%%%%%%%%%%%%%%%%%%%

\section{\label{formalism} Formalism}

Let us consider the following class of processes 
\begin{equation}
p_1 \; \to \; p_2 \;  f_1 \;\bar f_2\;,
\label{p1p2}
\end{equation}
where a fermion-antifermion pair $f_1$-$\bar f_2$  is emitted
as the particle $p_1$ is supposed to evolve into the particle $p_2$. 
The $f_1$, $\bar f_2$, $p_1$, and $p_2$ rest masses are $m_1$, $m_2$, 
$M_1$, and $M_2$,  respectively. We will be interested here in cases where 
$m_1, m_2 \ll M_1, M_2$. The fermion emission will be assumed not to 
change significantly the four-velocity of $p_2$ with respect to $p_1$. 
This is called {\em ``no-recoil condition''}, which is verified when the 
momentum of the emitted fermions (with respect to the inertial frame 
instantaneously at rest with particle $p_1$) satisfies 
$|{\bf k}| \ll M_1,\;M_2$. 
Because $m_1, m_2 \ll M_1, M_2$, this implies that the energy of each emitted
fermion satisfies
${\omega} \ll M_1,\;M_2$. As the typical energy $\omega$ 
of the emitted fermions  is of the order of the  proper acceleration 
$a$ of the particle $p_1$, the no-recoil condition can be recast 
as~\cite{VM3} 
\begin{equation}
a \ll M_1,\; M_2 \;.
\label{norecoilcondition}
\end{equation}

The particles $p_1$ and $p_2$ will be seen as distinct energy eigenstates
$\vert p_1 \rangle$ and $\vert p_2 \rangle$, respectively, of a two-level 
system. The associated {\it proper} Hamiltonian $\hat H_0$ of the 
particle system satisfies, thus, 
\begin{equation}
\hat H_0 \; \vert p_j \rangle = M_j \; \vert p_j \rangle\;,\;\; j=1,2\;.
\label{H0}
\end{equation}
Hence, we describe our pointlike particle system by the semiclassical 
(vector) current 
\begin{equation}
\hat j^\mu (x)=
\hat q(\tau)\;[u^\mu (\tau)/u^0 (\tau)] \; \delta^3 [{\bf x}-{\bf x}(\tau)]\;,
\label{CI}
\end{equation}
where $x^\mu(\tau)$ is the   classical world line associated with the particle 
system $p_1$-$p_2$, $u^\mu (\tau) \equiv dx^\mu/d\tau $ is the 
corresponding four-velocity, and 
$\hat q (\tau ) \equiv e^{i \hat H_0 \tau} \hat q_0 e^{-i \hat H_0 \tau}$, 
where $\hat q_0$ is a self-adjoint operator
evolved by the one-parameter group of unitary operators
$ \hat U (\tau ) = e^{-i\hat H_0 \tau}$. 

Each emitted fermion will be associated with a spinorial field written as
\begin{equation}
\hat \Psi(x)= \sum_{\sigma = \pm } \int d^3 {\bf k}
\left[ \hat b_{{\bf k} \sigma} \psi^{(+\omega)}_{{\bf k} \sigma} (x)
     + \hat d^\dagger_{{\bf k} \sigma} \psi^{(-\omega)}_{-{\bf k} -\sigma} (x) 
\right]\;,
\label{FF}
\end{equation}
where $ \hat b_{{\bf k} \sigma} $ and $ \hat d^\dagger_{{\bf k} \sigma}$ 
are annihilation and creation operators of fermions
and antifermions, respectively, with three-momentum ${\bf k}=(k^x,
k^y,k^z)$, energy $ \omega=\sqrt{{\bf k}^2+m^2}$ and polarization 
$\sigma$, and $ \psi^{(+\omega)}_{{\bf k} \sigma} $ and 
$ \psi^{(-\omega)}_{{\bf k} \sigma}$ are positive and negative 
frequency solutions of the Dirac equation
$i\gamma^\mu \partial_\mu \psi^{(\pm \omega)}_{{\bf k} \sigma} 
 - m \psi^{(\pm \omega)}_{{\bf k} \sigma} =0$.

Next, we minimally couple the spinorial fields $\hat \Psi_1$ 
and $\hat \Psi_2$ (associated with the two emitted fermions $f_1$-${\bar f}_2$,
respectively)
to our semiclassical current $\hat j^{\mu}$ (that describes the particle 
system $p_1$-$p_2$)
according to the 
weak-interaction action~\cite{IZ}-\cite{CQ}
\begin{eqnarray}
\hat S_I & = \int d^4x \;\hat j_\mu 
           \{\hat{\bar \Psi}_1 \gamma^\mu (c_V-c_A\gamma^5)\hat \Psi_2 
\nonumber \\
	 & +
        \hat{\bar \Psi}_2 \gamma^\mu (c_V-c_A\gamma^5)\hat \Psi_1 \} \;,
\label{S}
\end{eqnarray}
where $c_V=c_A=1$ in the cases here analyzed.

The  transition amplitude for the process (\ref{p1p2}) at the
tree level is given by
\begin{equation}
{\cal A}^{\sigma_1 \sigma_2}_{{\bf k}_1 {\bf k}_2} =
\; \langle  p_2 \vert \otimes \langle {f_1}_{{\bf k}_1 \sigma_1} , 
\bar {f_2}_{{\bf k}_2 \sigma_2} \vert \;
\hat S_I \;
\vert 0 \rangle \otimes \vert p_1  \rangle \; ,
\label{AMP}
\end{equation}
and the differential transition probability is
\begin{equation}
\frac{d{\cal P}^{p_1 \to p_2}}{d^3{\bf k}_1 d^3{\bf k}_2}\;=\;
\sum_{\sigma_1=\pm} \sum_{\sigma_2=\pm} \vert 
{\cal A}^{\sigma_1 \sigma_2}_{{\bf k}_1 {\bf k}_2} \vert^2 \;,
\label{dP1}
\end{equation}
which leads to (see Ref.~\cite{VM3} for details)
\begin{widetext}
\begin{eqnarray}
\frac{d{\cal P}^{p_1 \to p_2}}{d^3{\bf k}_1 d^3{\bf k}_2}
\;&=&\;
\frac{2 \; G_{\rm eff}^2}{(2\pi)^6 \omega_1 \omega_2}
\int_{-\infty}^{+\infty} d\tau 
\int_{-\infty}^{+\infty} d\tau'
e^{i\Delta M(\tau-\tau')}\;
e^{i(k_1+k_2)^\lambda [x(\tau)-x(\tau')]_\lambda}
\nonumber \\
&\times &
\left\{
\left[2 k_1^{(\mu} k_2^{\nu)} +
i \epsilon^{\mu\nu\alpha\beta}
k_{1\alpha} k_{2\beta} \right] u_\mu(\tau)u_\nu(\tau') 
- k_1^\alpha {k_2}_\alpha  u^\mu(\tau)u_\mu(\tau')
\right\}\; ,
\label{dP3}
\end{eqnarray}
\end{widetext}
where $\epsilon^{\mu\alpha\nu\beta}$ is the totally skew-symmetric 
Levi-Civita pseudo-tensor (with $\epsilon^{0123}=-1$),
$
k_1^{(\mu}k_2^{\nu)} \equiv 
(k_1^{\mu}k_2^{\nu}+k_1^{\nu}k_2^{\mu})/2,
$
$\Delta M \equiv M_2 - M_1$ 
and
$G_{\rm eff} \equiv \vert \langle p_2 \vert \hat q_0 \vert p_1 \rangle \vert$
is the effective coupling constant.

In those situations where the particle is accelerated by a background 
electromagnetic field, a full quantum-mechanical investigation  would be, 
{\em in principle}, 
possible. In this case, any recoil effects associated with the fermion 
emission would be  automatically taken into account. For instance, in 
Ref.~\cite{BLP} the {\em quasiclassical approach} was developed to consider 
$\gamma$-synchrotron radiation from an electron immersed in a classical  
background magnetic field with intensity
$H \ll H_0 = 4.4 \, \times \, 10^{13} $~Gauss 
with the electron Lorentz factor satisfying $\gamma \gg 1$. 
A similar approach was applied to the neutrino-antineutrino emission 
in Sec.~6.1 of Ref.~\cite{BKS}. Another very promising approach, which could 
be adapted to the present case, was recently developed by Higuchi who  
investigated the radiation reaction effect on accelerated
charges in the context of Quantum Field Theory~\cite{H}. In this vein, 
further developments of our  formalism to 
naturally take into account back-reaction effects would be welcome.
In spite of this, our semiclassical approach has the advantage
of being applicable to a quite general class of processes irrespective to
the acceleration source origin: electromagnetic, gravitational or
some other one. Moreover, it agrees with the full quantum  mechanical 
treatment used in the aforementioned cases
when the no-recoil condition is satisfied (i.e., $\chi \ll 1$ 
in Refs.~\cite{BLP}-\cite{BKS}). Hence, our approach and the
other ones in the literature ~\cite{BK}-\cite{H} should be seen
as complementing each other.

%%%%%%%%%%%%%%%%%%%%%%%%%%%%%%%%%%%%%%%%%%%%%%%%%%%%%%%%%%%
%!!!!!!!!!!!  UNIFORMLY SWIRLING  CURRENT   !!!!!!!!!!!!!!%
%%%%%%%%%%%%%%%%%%%%%%%%%%%%%%%%%%%%%%%%%%%%%%%%%%%%%%%%%%%

\section{\label{current} Uniformly swirling currents}

The world line of a  particle with uniform circular motion
with radius $R$ and angular velocity $\Omega$, as defined
by observers at rest in an inertial frame associated with 
inertial coordinates $(t,{\bf x})$, is
\begin{equation}
x^\mu(\tau)= (t\;,\;R \cos (\Omega t)\;,\;R \sin (\Omega t)\;,\; 0) \; ,
\label{WL}
\end{equation}
and the corresponding four-velocity is 
\begin{equation}
u^\mu(\tau)= \gamma \,
            (1\;,\; 
             - R \Omega \sin (\Omega t)\;,\;
             R \Omega \cos (\Omega t) \;,\;
             0)\;,
\label{UACC}
\end{equation}
where 
$\gamma \equiv (1-R^2 \Omega^2)^{-1/2} =  constant$ 
is the Lorentz factor ($v \equiv R \Omega < 1$),
$t=\gamma \tau$,
and  
$ a = \sqrt{-a_\mu a^\mu} = R \, \Omega^2 \gamma^2$
is the proper acceleration.

In order to decouple the integrals in Eq.~(\ref{dP3}),
we define new coordinates,
\begin{equation}
\sigma \equiv \gamma ({\tau-\tau '})/{2}\;\;\;\;{\rm and}\;\;\;\;
s \equiv \gamma({\tau+\tau '})/{2}\;,
\label{CCG}
\end{equation}
and perform the 
change in the momentum variable 
\begin{equation}
k^\mu \mapsto {\widetilde{k}}^\mu 
= 
( 
 {\widetilde{\omega}}\; , {\widetilde{\bf k}} 
)\;, 
\label{Gamma/k1k2}
\end{equation}
where
\begin{eqnarray}
{\widetilde{\omega}} \; &=& \omega \; , 
\nonumber \\
{\widetilde{k}^x}  &=&  k^x \cos (\Omega s) + k^y \sin (\Omega s) \;,
\nonumber \\
{\widetilde{k}^y}  &=&  - k^x \sin (\Omega s) + k^y\cos (\Omega s) \;,
\nonumber \\
{\widetilde{k}^z}  &=&  k^z \; ,
\nonumber 
\end{eqnarray}
which consists of a rotation by an angle $\Omega s$ around 
the $k^z$ axis. Hence, we obtain from Eq.~(\ref{dP3}) the  
following transition rate per momentum-space element of each
emitted fermion:
\begin{widetext}
\begin{eqnarray}
&& \frac{ 
   d\Gamma^{p_1 \to p_2}}{d^3{\widetilde{\bf k}}_1 d^3{\widetilde{\bf k}}_2
     }
 \; = \;
\frac{
 2\;\gamma\;G_{\rm eff}^2}{(2\pi)^6{\widetilde{\omega}}_1{\widetilde{\omega}}_2
     }
\int_{-\infty}^{+\infty} d\sigma\;
\exp \left[ 
        i\left(
          \Delta M \sigma/\gamma  + 
          (\widetilde{k}_1 + \widetilde{k}_2)^\mu X_\mu (\sigma) 
         \right) 
     \right]
\nonumber \\
&&
\times \left[
(
 {\widetilde{\omega}}_1 {\widetilde{\omega}}_2 + 
 {\widetilde{\bf k}}_1 \cdot {\widetilde{\bf k}}_2
)
- R^2 \Omega^2 
(
 {\widetilde{k}}^x_1 {\widetilde{k}}^x_2 - 
 {\widetilde{k}}^y_1 {\widetilde{k}}^y_2
)
+\, R^2 \, \Omega^2 
(
 {\widetilde{\omega}}_1 {\widetilde{\omega}}_2 -
 {\widetilde{k}}^z_1 {\widetilde{k}}^z_2 
)
\cos (\, \Omega \sigma\,  )
\right.
\nonumber \\
& &
-\,  2 \, R \, \Omega \, 
(
{\widetilde{\omega}}_1 {\widetilde{k}}^y_2 +
{\widetilde{\omega}}_2 {\widetilde{k}}^y_1
)
\cos (\, \Omega \sigma/2\,  ) 
+ \, 2\,  i\,  R \, \Omega \, 
(
{\widetilde{\bf k}_1} \times {\widetilde{\bf k}_2}
)^x
\sin (\, \Omega  \sigma/2 \, )
\nonumber \\
& &
\left.
- \, i \, R^2 \, \Omega^2\, 
(
{\widetilde{\omega}}_1 {\widetilde{k}}^z_2 -
{\widetilde{\omega}}_2 {\widetilde{k}}^z_1
)
\sin (\, \Omega \sigma \, )
\right] \; ,
\label{dG1}
\end{eqnarray}
\end{widetext}
where 
$ 
\Gamma^{p_1 \to p_2} \equiv \gamma\; {d{\cal P}^{p_1 \to p_2}}/{ds} \; 
$ 
is the transition probability per {\em proper} time and
$$
X^\mu (\sigma) \equiv 
(\sigma\, ,\, 0\, ,\, 2R\sin (\Omega \sigma/2) \, ,\, 0) \;.
$$

In order to calculate the transition rate 
\begin{equation}
\Gamma^{p_1 \to p_2} 
\equiv 
\int d^3 \widetilde{\bf k}_1
\int d^3 \widetilde{\bf k}_2
\frac{d\Gamma^{p_1 \to p_2}}{d^3 \widetilde{\bf k}_1 d^3 \widetilde{\bf k}_2}\;,
\label{Gaux}
\end{equation}
it is convenient to use Eq.~(\ref{dG1}) to rewrite Eq.~(\ref{Gaux}) as
\begin{equation} 
\Gamma^{p_1 \to p_2}
=
\frac{2\; \gamma\; G_{\rm eff}^2}{(2\pi)^6 }
\int_{- \infty}^{+\infty} d \sigma \;
e^{i\; \Delta M \; \sigma/\gamma} 
G_{\mu \nu} A^{\mu \nu} \; ,
\label{Gaux2}
\end{equation}
where
\begin{equation}
G_{\mu \nu} 
\equiv 
- \frac{\partial I_1}{\partial X^\mu} 
  \frac{\partial I_2}{\partial X^\nu}
\label{G}
\end{equation}
with
\begin{equation}
I_l  
\equiv 
\int d^3 \widetilde{\bf k}_l
\frac{e^{i\; \widetilde{\bf k}_l^\lambda X_\lambda}}
{\widetilde\omega_l} 
\;, \;\;\;\; l=1,2\;,
\label{I_a}
\end{equation}
and $\widetilde{\omega}_l = \sqrt{\widetilde{\bf k}_l^2 + m_l^2}$, and
\begin{widetext}
\begin{equation}
A_{\mu \nu} =
\left[
 \begin{array}{cccc}
 1+R^2\Omega^2\cos(\Omega\sigma) & 0 & 
             -2R\Omega\cos(\Omega\sigma/2) & -iR^2\Omega^2\sin(\Omega\sigma)\\
 0 & 1-R^2\Omega^2 & 
             0 & 0\\
 -2R\Omega \cos (\Omega\sigma/2) & 0 & 
             1+R^2\Omega^2 & 2iR\Omega \sin(\Omega\sigma/2)\\
 iR^2\Omega^2\sin(\Omega\sigma) & 0 & 
            -2iR\Omega\sin(\Omega\sigma/2) & 1-R^2\Omega^2\cos(\Omega\sigma)    
 \end{array} 
\right] \; .
\label{A}
\end{equation}
\end{widetext}
In order to integrate Eq.~(\ref{I_a}), we introduce 
spherical coordinates in the momenta space 
$
({\widetilde{k}}_l\in {\rm R}^+,
 {\widetilde{\theta}}_l\in [0,\pi],
 {\widetilde{\phi}}_l\in [0,2\pi))
$, 
where
$
{\widetilde{k}}_l^x 
= 
{\widetilde{k}}_l \sin {\widetilde{\theta}_l} \cos {\widetilde{\phi}_l}
$,
$
{\widetilde{k}}_l^y 
= 
{\widetilde{k}}_l \sin {\widetilde{\theta}_l} \sin {\widetilde{\phi}_l}
$,
and
$
{\widetilde{k}}_l^z 
=
{\widetilde{k}}_l \cos {\widetilde{\theta}_l} 
$.
By doing so, we obtain
$$
I_l  = \frac{4 \pi}{|{\bf X}|} 
       \int_{m_l}^{+\infty} d{\widetilde{\omega}_l} 
       e^{i{\widetilde{\omega}_l} X^0 }
          \sin
          \left[ 
          \sqrt{{\widetilde{\omega}_l}^2 - m_l^2} \; | {\bf X }| 
          \right]\; ,
$$
where $|{\bf X}| \equiv \sqrt{-X_i X^i\,} $.
Next, by redefining the frequency  variable as
$\widetilde{\omega}_l \equiv m_l \cosh \xi $, 
we obtain
$$
I_l  = \frac{-2\pi i m_l}{ |{\bf X}| }
\int_{-\infty}^{+\infty} d\xi 
e^{i m_l (X^0 \cosh \xi + | {\bf X} | \sinh \xi) } \sinh \xi .
$$
Now, we perform the change of  variable
$\xi \mapsto \eta \equiv e^{\xi}$, leading to
\begin{eqnarray}
I_l = 
\frac{ i \pi m_l}{| {\bf Y} |}
\int_0^{+\infty} 
& &
d\eta (\eta^{-2} -1)
\exp 
\left[
       \frac{i m_l (Y^0 + | {\bf Y} |) \eta}{2} 
\right.       
\nonumber \\
& &
\left. 
     + \frac{i m_l (Y^0 - | {\bf Y} |)}{2 \eta}
     \right] \;,
\end{eqnarray}
where we have introduced a small positive regulator $\epsilon > 0$
in the integral as follows:
$$
X^\mu \mapsto Y^\mu = (X^0 + i\epsilon, X^1, X^2, X^3)\; .
$$
(Note that $ |{\rm Re} (Y^0)| = |X^0| > |{\bf X}| = | {\bf Y} | $.)
Then, by using expressions (3.471.11) and (8.484.1) of Ref.~\cite{GR}, 
we obtain
\begin{equation}
I_l 
 = 
\frac{-2 \, \pi^2 \, i\, \, m_l {\rm sign}(\sigma)  }{\sqrt{Y_\mu Y^\mu \,}}
H_1^{(1)} \left( {\rm sign}(\sigma) m_l \sqrt{Y_\mu Y^\mu \,} \right) \;,
\label{I_afim}
\end{equation}
where
$H^{(1)}_\mu (z)$ is the Hankel function of the first kind.
As a result, by making the variable change 
$\sigma \mapsto \lambda \equiv - a \sigma/\gamma$
and by defining $Z^\mu \equiv (a/\gamma) Y^\mu $,
the transition rate~(\ref{Gaux2}) can be cast in the form
(see also expression 8.472.4 in Ref.~\cite{GR})
\begin{widetext}
\begin{equation} 
\Gamma^{p_1 \to p_2}
= 
\frac{G_{\rm eff}^2 \widetilde{m}_1^4 \widetilde{m}_2^4 a^5\gamma^4}{8 \pi^2 }
\int_{- \infty}^{+\infty} d \lambda  \;
e^{- i\; \widetilde{\Delta M}\, \lambda } 
Z^\mu Z^\nu A_{\mu \nu} 
\frac{ H_2^{(1)} ( z_1 ) }{z_1^2}\; 
\frac{ H_2^{(1)} ( z_2 ) }{z_2^2}  \; ,
\label{Gfim}
\end{equation}
where we have defined
$\widetilde{m_l} \equiv m_l/a$,
$\widetilde{\Delta M} \equiv \Delta M/a$,
$\epsilon'\equiv a \epsilon/\gamma \ll 1$,
$z_l \equiv - \widetilde{m}_l \gamma {\rm sign}(\lambda) 
\sqrt{Z_\lambda Z^\lambda \,}$,
and where
$
Z^\mu
       = (-\lambda + i \epsilon', 
          0, 
          -(2Ra/\gamma) \sin (\Omega \lambda \gamma/2a) ,
          0) 
$
with
\begin{equation}
A_{\mu \nu} =
\left[
 \begin{array}{cccc}
 1+R^2\Omega^2\cos(\Omega\gamma\lambda/a) & 0 & 
                              -2R\Omega\cos(\Omega\gamma\lambda/2a) & 
                                     iR^2\Omega^2\sin(\Omega\gamma\lambda/a)\\
 0 & 1-R^2\Omega^2 & 0 & 0\\
 -2R\Omega \cos (\Omega \gamma \lambda/2 a) & 0 &  
                              1+R^2\Omega^2 & 
                                     -2iR\Omega\sin[\Omega\gamma\lambda/2a]\\
 -iR^2\Omega^2\sin(\Omega\gamma\lambda/a) & 0 & 
                              2iR\Omega\sin[\Omega\gamma\lambda/2a] &   
                                     1-R^2\Omega^2 \cos(\Omega \gamma \lambda/a)
 \end{array} 
\right] \; .
\label{Afin}
\end{equation}
\end{widetext}
It is not easy to integrate Eq.~(\ref{Gfim}) in general. Notwithstanding,
we will be interested in the physically relevant regime where
$\widetilde{m}_l \ll 1$. In this limit, Eq.~(\ref{Gfim}) can be cast in a more 
suitable form by using the following expansion for the Hankel 
function~\cite{GR}:
\begin{equation}
H_2^{(1)} (z_l) \approx -\frac{4 i }{\pi z_l^2} -\frac{i}{\pi}
                         + {\cal O} (z_l^2 \ln z_l) 
\; \; {\rm for} \;\; |z_l| \ll 1 \; .
\label{approx1}
\end{equation}
We note that for $|\lambda |$ large enough, $|z_l|>1$, in which case
the expansion~(\ref{approx1}) ceases to be a good approximation. 
[For instance, for $\gamma^2 \gg 1/\widetilde{m}_l \gg 1$, we have that
$|z_l|>1$ for $|\lambda | \geq 1/\sqrt{12 \widetilde{m}_l \,}$ ($l=1,2$),
while for $1/\widetilde{m}_l \gg \gamma^2 \gg 1$, we have that 
$|z_l|>1$ for $|\lambda | \geq 1/ (\gamma \widetilde{m}_l )$.]
Notwithstanding, this will not be important
because the error committed in this region will be small to affect 
the final result provided that $\widetilde{m}_l \ll 1$. Hence we write 
Eq.~(\ref{Gfim}) in the form
\begin{eqnarray} 
 \Gamma^{p_1 \to p_2}
 \approx & &
\frac{-G_{\rm eff}^2  a^5 }{8 \pi^4}
\int_{- \infty}^{+\infty} d \lambda \;
e^{- i\; \widetilde{\Delta M}\, \lambda } 
\frac{Z^\mu Z^\nu A_{(\mu \nu)}}{(Z^\lambda Z_\lambda)^2}
\nonumber \\
& &
\times 
\left(
 \frac{16 }{\gamma^4 (Z_\lambda Z^\lambda)^2}
+\frac{4  (\widetilde{m}_1^2 + \widetilde{m}_2^2)}{\gamma^2 Z_\lambda Z^\lambda}
\right)\; ,
\label{Gfimaprox1}
\end{eqnarray}
where
\begin{equation}
Z_\lambda Z^\lambda  = 
(\lambda-i\epsilon')^2 - (2 R a/\gamma)^2 \sin^2 (\Omega \lambda \gamma/2a)\,.
\label{W}
\end{equation}
Eventually, Eq.~(\ref{Gfimaprox1}) can be seen as the expansion of the
reaction rate up to second order in $\widetilde{m}_l \ll 1$.
In order to solve this integral, we expand  $Z^\lambda Z_\lambda$
for relativistic swirling particles~\cite{LPf}-\cite{T}, i.e., $\gamma \gg 1$
(recall that $R= v^2 \gamma^2/ a$, $\Omega = a/(v \gamma^2)$,
and $v=\sqrt{1-\gamma^{-2}}$): 
\begin{eqnarray}
Z_\lambda Z^\lambda  
& \approx &
\frac{1}{12\, \gamma^{2}} 
(\lambda + i \sqrt{3} A_+)(\lambda + i \sqrt{3} A_-)
(\lambda - i \sqrt{3} B_+)
\nonumber \\
& &\times 
(\lambda - i \sqrt{3} B_-) \; ,
\end{eqnarray} 
where
$$
A\mp \equiv 1 \mp \sqrt{1+ 2 \widetilde{\epsilon} /\sqrt{3}} 
$$
and
$$
B\mp \equiv 1 \mp \sqrt{1- 2 \widetilde{\epsilon} /\sqrt{3}}
$$
with $\widetilde{\epsilon} \ll 1$. 
For $|\lambda| \gtrsim 2 v \gamma$, where the 
expansion ceases to be a good approximation, the integral 
contributes very little again and, thus, will not have any major 
influence in the final result. 
Thus, the integral in Eq.~(\ref{Gfimaprox1}) can be rewritten in the complex
plane:
\begin{eqnarray} 
\Gamma^{p_1 \to p_2}
\approx & &
\frac{-G_{\rm eff}^2  a^5 }{8 \pi^4}
\oint_{C} d \lambda \;
e^{- i\; \widetilde{\Delta M}\, \lambda  } 
\frac{Z^\mu Z^\nu A_{(\mu \nu)}}{(Z^\lambda Z_\lambda)^2}
\nonumber \\
& &
\times
\left(
 \frac{16}{\gamma^4 (Z_\lambda Z^\lambda)^2}
+\frac{4  (\widetilde{m}_1^2 + \widetilde{m}_2^2)}{\gamma^2 Z_\lambda Z^\lambda}
\right)\; ,
\label{Gfimaprox2}
\end{eqnarray}
where the complex integration path,
given by 
$
C \equiv (-L,L) \cup \{L \; e^{i \theta}, \theta \in [-\pi, 0] \}
$
with $L\to \infty$, is  counter-clockwise oriented. This expression, then,
can be performed by using residues.
We present below the results obtained for the leading term 
in $\gamma$ \cite{software}:
\begin{widetext}
\begin{eqnarray}
&& \Gamma^{p_1 \to p_2}
\approx
\frac{
      G_{\rm eff}^2  a^{5} \exp({- 2\sqrt{3}\widetilde{\Delta M}})
     }{
      1728 {\pi }^3
      }
\left( 49 \sqrt{3} + 102\widetilde{\Delta M} 
      + 30\sqrt{3}\widetilde{\Delta M}^2
      + 12\,\widetilde{\Delta M}^3       
\right.
\nonumber \\
& &         - 39\,\sqrt{3}\,( {\widetilde{m}_1}^2 + {\widetilde{m}_2}^2 )
            - 90\,\widetilde{\Delta M}\, 
               ( {\widetilde{m}_1}^2 + {\widetilde{m}_2}^2)
\left.      
      - 36\,\sqrt{3}\,\widetilde{\Delta M}^2\,({\widetilde{m}_1}^2
      + {\widetilde{m}_2}^2) 
\right) \;,
%\label{taxadecaimentoinercial}
\end{eqnarray}
\end{widetext}
where we recall that this is valid for $\widetilde{m}_1,\widetilde{m}_2 \ll 1$
and $\gamma \gg 1$.
 
Next, we calculate the radiated power in form of each fermion,
\begin{equation}
W^{p_1 \to p_2}_{l} 
\equiv 
\int d^3 \widetilde{\bf k}_1
\int d^3 \widetilde{\bf k}_2
\; \widetilde{\omega}_l
\frac{
      d \Gamma^{p_1 \to p_2}
     }{
     d^3 \widetilde{\bf k}_1 d^3 \widetilde{\bf k}_2
     }\;, 
\label{Waux}
\end{equation}
where the  index $l=1,2$
is used to distinguish which fermion we are referring to.
We write Eq.~(\ref{Waux}) as
\begin{equation} 
W^{p_1 \to p_2}_{1}
=
\frac{2\; G_{\rm eff}^2}{(2\pi)^6 }
\int_{- \infty}^{+\infty} d \sigma \;
e^{i\; \Delta M \; \sigma/\gamma} 
H_{\mu \nu} A^{\mu \nu} \; ,
\label{Waux2}
\end{equation}
where we have chosen (with no loss of generality) $l=1$, 
i.e., we are computing the radiated power associated with the
fermion with mass $m_1$. Here
\begin{equation}
H_{\mu \nu} 
\equiv 
- \frac{\partial J_1}{\partial X^\mu} 
  \frac{\partial I_2}{\partial X^\nu} \;,
\label{H}
\end{equation}
\begin{equation}
J_1  
\equiv 
\int d^3 \widetilde{\bf k}_1 
e^{i\; \widetilde{\bf k}_1^\lambda X_\lambda} \;,
\label{J_1}
\end{equation}
and $I_2$ is given by Eq.~(\ref{I_a}).
By following the same steps used to integrate $I_l$, which  allowed us to 
reach Eq.~(\ref{I_afim}), we obtain
\begin{equation}
J_1 = 
\frac{2 \, \pi^2  \, m_1^2\, Y_0  }{Y_\mu Y^\mu }
H_2^{(1)} \left( {\rm sign}(\sigma) m_1 \sqrt{Y_\mu Y^\mu \,} \right) \;.
\label{J_1fim}
\end{equation}
As a result, by making again the variable change 
$\sigma \mapsto \lambda \equiv - a \sigma/\gamma$
and by defining $Z^\mu \equiv (a/\gamma) Y^\mu $,
the emitted power~(\ref{Waux2}) can be cast in the form
(see also expression 8.472.4 in Ref.~\cite{GR})
\begin{widetext}
\begin{eqnarray} 
& & W^{p_1 \to p_2}_{1}
=
\frac{
      G_{\rm eff}^2 \widetilde{m}_1^4 \widetilde{m}_2^4 \gamma^2 a^6 \,i
      }{
      8 \pi^2
      }
\int_{- \infty}^{+\infty} d \lambda \;
e^{- i\; \widetilde{\Delta M}\, \lambda  } 
\frac{ H_2^{(1)} ( z_2 ) }{z_2^2}
\nonumber \\
&& \times
\left[ \frac{ H_3^{(1)} ( z_1 ) }{z_1^3}\;
      \widetilde{m}_1^2 \gamma^2 Z^0Z^\mu Z^\nu A_{(\mu \nu)}
      -\frac{ H_2^{(1)} ( z_1 ) }{z_1^2}\; \eta^{0 \mu} Z^\nu A_{\mu \nu}      
\right] \;,
\label{Wfim}
\end{eqnarray}
\end{widetext}
where we recall that $z_l$, $Z^\mu$ and $A_{\mu \nu}$ are the same ones 
defined below Eq.~(\ref{Gfim}).
In order to perform this integral in the limit $\widetilde{m}_l \ll 1$, 
we use the approximation~(\ref{approx1}) and (see Ref.~\cite{GR})
\begin{equation}
H_3^{(1)}(z_l)\approx -\frac{16 i }{\pi z_l^3} -\frac{2i}{\pi z_l}
                      -\frac{z_l i}{4 \pi} 
                      + {\cal O} (z_l^3 \ln z_l) 
\label{approx2}
\end{equation} 
for $ |z_l| \ll 1 $. Then, by letting Eqs.~(\ref{approx1}) and 
(\ref{approx2}) in Eq.~(\ref{Wfim}), we can perform the remaining 
integral in the complex plane along the path
$
C \equiv (-L,L) \cup \{L \; e^{i \theta}, \theta \in [-\pi, 0] \}
$
with $L\to \infty$, as for the reaction rate, and obtain 
the emitted power $W^{p_1 \to p_2}_{1}$.
We present below the result for the leading term in 
$\gamma$~(see Ref.~\cite{software}):
\begin{widetext}
\begin{eqnarray} 
&&
\!\!\!\!
W^{p_1 \to p_2}_{1}
\!\approx\!
\frac{G_{\rm eff}^2 a^6 e^{-2\sqrt{3}\; \widetilde{\Delta M}}}{3456 \pi^3 }
\left[
      320 
      + 241\sqrt{3} \; \widetilde{\Delta M} 
      + 246 \widetilde{\Delta M}^2 
      + 46\sqrt{3}\;\widetilde{\Delta M}^3
      + 12 \widetilde{\Delta M}^4  
      - 48 ( \widetilde{m}_1^2  
\right.
\nonumber \\      
&&
\!\!\!
\left.
  + 5\widetilde{m}_2^2 ) 
  - 3 \sqrt{3} \widetilde{\Delta M} (17 \widetilde{m}_1^2 + 65 \widetilde{m}_2^2)
  - 18 \widetilde{\Delta M}^2 ( 5 \widetilde{m}_1^2 + 13 \widetilde{m}_2^2 )
  - 24 \sqrt{3} \widetilde{\Delta M}^3 (\widetilde{m}_1^2+ 2 \widetilde{m}_2^2)
\right]
\label{w1}
\end{eqnarray}
\end{widetext}
where we recall that this is valid for $\widetilde{m}_1,\widetilde{m}_2 \ll 1$
and $\gamma \gg 1$. Clearly, $W^{p_1 \to p_2}_{2}$ is obtained by permuting
$m_1\longleftrightarrow m_2$ in Eq.~(\ref{w1}).
\setcounter{equation}{0}
%%%%%%%%%%%%%%%%%%%%%%%%%%%%%%%%%%%%%%%%%%%%%%%%%%%%%%%%%%%
%!!!!!!!!!!!!!!!!!    PROTON  DECAY   !!!!!!!!!!!!!!!!!!!!%
%%%%%%%%%%%%%%%%%%%%%%%%%%%%%%%%%%%%%%%%%%%%%%%%%%%%%%%%%%%

\section{\label{proton} Proton decay}

Seemingly, Ginzburg and Syrovatskii~\cite{GS} were the first ones 
to comment about the decay of noninertial protons, but only recently 
Muller~\cite{M} presented the first estimative for the decay rate
of the inverse $\beta$-decay  
\begin{equation}
p \to n \; e^+ \; \nu  
\label{pn}
\end{equation}
by assuming that all the particles were scalars. Further, the authors 
used the semiclassical approach (where the leptons are described by 
fermionic fields indeed) to calculate the decay rate for uniformly accelerated 
protons. Here we analyze the case of swirling
protons, which can model high-energy protons moving in the 
magnetosphere of a pulsar. 
\begin{figure}
\epsfig{file=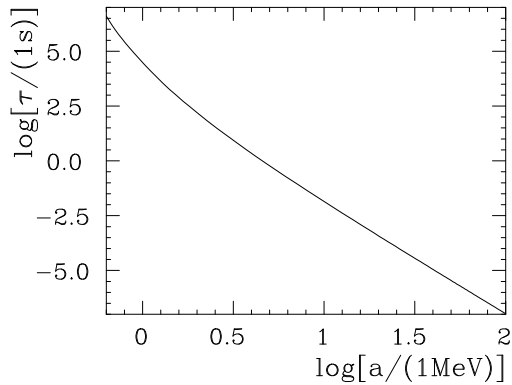,angle=90,width=0.9\linewidth,clip=}
\caption{\label{tempofracog100}
 The proton mean proper lifetime $\tau$ 
is plotted as a function of its proper acceleration $a$,
where we have assumed $\gamma = 100$. 
The result is not a very sensitive function of $\gamma$ 
provided that $\gamma \gg 1$. $\tau \propto 1/a^5$ 
for sufficiently large $a$.}
\end{figure}

The effective coupling constant  for the inverse $\beta$-decay, 
$G_{\rm eff}= G_{pn}$, is obtained by imposing that 
the mean proper lifetime of inertial 
{\em neutrons} be $887$ s~\cite{PDG}, i.e.,
\begin{equation}
\Gamma^{n\to p}_{in} \equiv
\Gamma^{n\to p}(\Omega\to 0) 
= \hbar / (887 \; {\rm s} )\;.
\label{GEFF}
\end{equation}
Of course, we cannot use our expression~(\ref{Gfimaprox2}) in this
case since it is not valid when $a < m_e$. Fortunately, however,
$\Gamma^{n \to p}$  can be integrated for inertial neutrons directly
from Eq.~(\ref{Gaux}) by making $\Omega = 0$ in
Eq.~(\ref{dG1}). This is achieved by a change of the momentum variables
as shown in Eq.~(\ref{Gamma/k1k2}). After performing the 
corresponding integrations in the angular coordinates and in 
$\widetilde{\omega}_e$, we obtain   
\begin{eqnarray}
\Gamma^{n\to p}_{in} & = &
\frac{G_{pn}^2 }{\pi^3}
\int_0^{\Delta M-m_e} d{\widetilde{\omega}}_\nu\;{\widetilde{\omega}}_\nu^2
\left(
\Delta M  - \widetilde{\omega}_\nu
\right)
\nonumber \\
& \times &
\sqrt{
\left(
\Delta M - \widetilde{\omega}_\nu 
\right)^2- m_e^2}\;,
\label{dGIN2}
\end{eqnarray}
where we have assumed $m_\nu = 0$.
By evaluating numerically Eq.~(\ref{dGIN2})  with 
$m_e= 0.511 \; {\rm MeV}$ and
$\Delta M= (m_n-m_p)= 1.29 \;{\rm MeV}$, we end up with
$
\Gamma^{n\to p}_{in}\;=\;
1.81 \times 10^{-3} \;G_{pn}^2 \; {\rm MeV}^5
$.
As a result, in order to fit Eq.~(\ref{GEFF}), 
we must set  
$
G_{pn}\;=\;1.74\;G_F\;,
$
where $G_F \equiv 1.166\times 10^{-5}\;{\rm GeV}^{-2}$
is the Fermi coupling constant~\cite{PDG}.
This phenomenological procedure has the advantage of by passing any 
uncertainties on the influence of the nucleon inner structure.
\begin{figure}
\epsfig{file=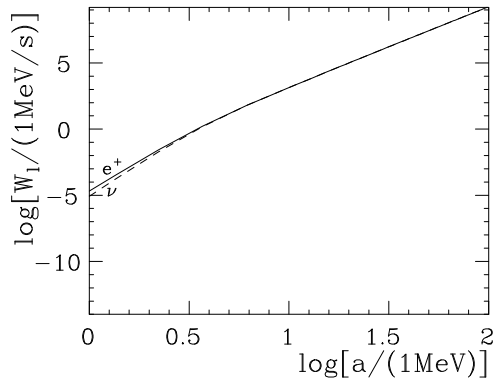,angle=90,width=0.9\linewidth,clip=}
\caption{\label{potenciafracaenug100}
 $W_{e^+}$ and $W_{\nu}$ are plotted as functions of the proton 
proper acceleration with solid and dashed lines, respectively. Although 
we have assumed $\gamma= 100$ in the  numerical calculation, the result 
is not a very sensitive function of $\gamma$ provided that $\gamma \gg 1$.}
\end{figure}

Now we are able to use Eq.~(\ref{Gfimaprox2}) to plot
the proton mean {\em proper} lifetime $\tau (a)= 1/\Gamma^{p\to n} $ 
(see Fig.~\ref{tempofracog100}). 
We have plotted the {\em proper} lifetime $ \tau (a)$ rather than the 
{\em laboratory} lifetime $t(a)$ in order to make it easier the
comparison  of this figure with  Fig.~1 in Ref.~\cite{VM3}.
[We have only considered 
accelerations $a \ll m_p=938$ MeV in order to respect our no-recoil 
condition (\ref{norecoilcondition}).] We notice 
that swirling protons decay somewhat faster than uniformly accelerated 
protons with the same proper acceleration $a$.  
We also exhibit how much energy is carried out in form of electrons
and neutrinos as calculated in Sec.~\ref{formalism}, by plotting
the emitted powers  $W_{l}$ for $l=e^+, \nu$ in 
Fig.~\ref{potenciafracaenug100}.

Astrophysics seems to provide suitable conditions for the 
observation of the decay of accelerated protons. 
Let us consider a cosmic ray proton
with energy $E_p = \gamma m_p \approx 1.6 \times 10^{14}$ eV 
under the influence of a magnetic field $B \approx 10^{14}$ Gauss
of a pulsar.  Protons under these
conditions have proper accelerations of $a_B = \gamma
e B/m_p \approx 110$ MeV $\gg m_e$. 
For these values of $E_p$ and $B$, the proton is confined
in a cylinder with typical radius 
$R \approx \gamma^2/a_B \approx 5\,\times \, 10^{-3} \; {\rm cm}
\ll l_B , $ where $l_B$ is the typical size of the magnetic field 
region. By using Eq.~(\ref{Gfimaprox2}), we obtain that such  protons 
would have a ``laboratory'' mean lifetime of
$t_p = \gamma \tau \approx 1.2 \; \times \; 10^{-2}$ s. 
Thus, under such conditions, protons rapidly decay.
For $l_B \approx 10^7$ cm, we  obtain that about 
$| \Delta N_p /N_p | \approx (1-e^{-l_B/t_p}) \approx 2.7\%$
of a bunch of protons would decay in this way.  
Hence our original estimative achieved by assuming
uniformly accelerated protons was roughly correct but still
2.7 times smaller than this more precise value.  We note that we did 
not take into account the
influence of the magnetic field on the emitted positron.
Clearly a more precise estimative should take into account
this effect as well as other ones as, e.g., the
non-uniformity of the magnetic field and energy losses
through electromagnetic synchrotron radiation.  The latter,
in particular, may not be a problem since extra energy may be
furnished to the proton from dynamo processes.  A
comprehensive analysis of such astrophysical issues will be
discussed elsewhere.

%%%%%%%%%%%%%%%%%%%%%%%%%%%%%%%%%%%%%%%%%%%%%%%%%%%%%%%%%%%%%%%%%%%%
%%%%%%%%%%%%%%%%%%      Neutrino Emission      %%%%%%%%%%%%%%%%%%%%%
%%%%%%%%%%%%%%%%%%%%%%%%%%%%%%%%%%%%%%%%%%%%%%%%%%%%%%%%%%%%%%%%%%%%

\section{\label{neutrino} Neutrino emission from uniformly swirling electrons}

Let us, now, consider the emission of
neutrino- antineutrino pairs from accelerated electrons,
\begin{equation}
e^- \to e^- \; \nu_e \; \bar \nu_e\;,
\label{ee}
\end{equation}
and compare our results in the proper limit with the ones in the 
literature obtained in the particular case where the electrons 
are quantized in a background magnetic field~\cite{BK}-\cite{LP}. 
\begin{figure}
\epsfig{file=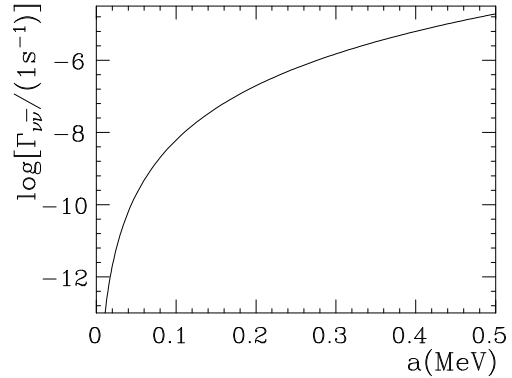,angle=90,width=0.9\linewidth,clip=}
\caption{\label{taxafracag100}
 The emission rate of $\nu_e$-${\bar \nu}_e$ 
pairs is plotted  for $a\leq m_e$ and $\gamma=100$. 
}
\end{figure}

The emission rate and  the  {\it total} radiated power 
of neutrino-antineutrino pairs can be calculated 
from the Sec.~IV results by assuming $ \Delta M =m_\nu =0$:
\begin{equation}
\Gamma_{\nu {\bar \nu}}=
\frac{\sqrt{3\,} \; G_{e\nu}^2\;a^5 }{3458 \pi^3} 
\left( 98 + 31/\gamma^2\right) + {\cal O} (\gamma^{-4})\;
\label{GNU}
\end{equation}
and
\begin{equation}
{\cal W}_{\nu {\bar \nu}}=
\frac{G_{e\nu}^2\;a^6}{135 \pi^3} 
\left( 
25 + 7/\gamma^2
\right)
+
{\cal O} (\gamma^{-4})
\;,
\label{WNU}
\end{equation}
where $G_{e\nu}$ is the corresponding effective coupling constant.

In order to determine the value of $G_{e\nu}$, we 
compare Eq.~(\ref{WNU}) with the neutrino-antineutrino 
radiated power obtained in the particular case where the 
electron is uniformly swirling in a constant magnetic field 
$B$, provided that its proper acceleration 
$a=\gamma e B/m_e \ll m_e$ (no-recoil condition). 
This can be easily 
calculated from the differential emission rate 
given, e.g., in Ref.~\cite{LP} or Ref.~\cite{BKS}  
(see Eq.~(6.6) in Ref.~\cite{VM3} for the final 
result below),
\begin{equation} 
{W}^{LP}_{\nu {\bar \nu}} 
=
\frac{5\;(2\;C_V^2+23\;C_A^2)}{108\pi^3}\;
G_F^2\;m_e^6\chi^6\;,
\label{W2}
\end{equation} 
where the vector and axial contributions to the
electric current are $C_V^2=0.93$ and $C_A^2=0.25$ \cite{KLYAH},
respectively, and $\chi \equiv a/m_e \ll 1$.
Thus, by comparing
$
{\cal W}^{LP}_{\nu {\bar \nu}} =
1.14\times 10^{-2}\;G_F^2\;a^6
$
with our Eq.~(\ref{WNU}), we obtain
$
G_{e\nu} = 1.38 \;G_F,
$
which is $40 \%$ smaller than the one obtained with our original
estimative with uniformly accelerated electrons.
\begin{figure}
\epsfig{file=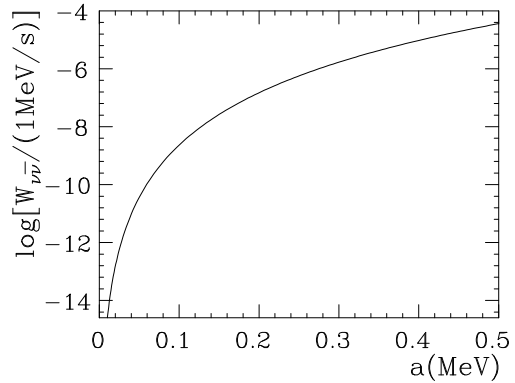,angle=90,width=0.9\linewidth,clip=}
\caption{\label{potenciafracanunug100}
The total $\nu_e$-${\bar \nu}_e$ radiated power is plotted for
$a \leq m_e$ and $\gamma = 100$. 
}
\end{figure}
In 
Figs.~\ref{taxafracag100}  and \ref{potenciafracanunug100}
we plot Eqs.~(\ref{GNU}) and (\ref{WNU}), respectively, for 
swirling electrons with $a\leq m_e$ and $\gamma = 100$.
We note that for the same electron proper acceleration, the 
neutrino-antineutrino emission rate is somewhat smaller than the 
one obtained for uniformly accelerated electrons.

%%%%%%%%%%%%%%%%%%%%%%%%%%%%%%%%%%%%%%%%%%%%%%%%%%%%%%%%%%%%%%%%%%%%
%%%%%%%%%%%%%%%%%%           Discussion        %%%%%%%%%%%%%%%%%%%%%
%%%%%%%%%%%%%%%%%%%%%%%%%%%%%%%%%%%%%%%%%%%%%%%%%%%%%%%%%%%%%%%%%%%%

\section{\label{discussion} Discussion}

We have investigated the weak-interaction emission of spin-$1/2$
fermions from decaying (and non-decaying) uniformly swirling particles.
As a particular application, we have focused on the inverse
$\beta$-decay of  uniformly swirling protons.
We have shown that high-energy protons in background magnetic
fields may have a considerably short lifetime.
By restricting our semiclassical current to behave classically,
i.e., by making $\Delta M \to 0$, we were able to use our formalism to
investigate the neutrino-antineutrino pair emission from uniformly
accelerated electrons and compare our results with the
ones in the literature obtained by quantizing the
electron field in a background magnetic field. By comparing
the results obtained for uniformly {\em accelerated} and {\em swirling}
particles, we conclude that depending on the accuracy level
required, one can use directly the formulas derived for uniformly
accelerated currents to make a reasonable estimative for
reaction rates and emitted powers associated with processes involving
accelerated particles as the ones treated here. This may be particularly
useful in some astrophysical situations.

Finally, it is worth mentioning that the approach of coupling a two-level
system to quantized fields in order
to describe the decay of accelerated particles is by no means restricted to
weak-interaction processes. In fact, the decaying of accelerated protons through
\begin{equation}
p \to n \; \pi^+
\label{strong}
\end{equation}
can be also analyzed in this framework by coupling a two level (scalar)
system to a massive Klein-Gordon quantum field describing the pion. Because
this is a strong-interaction process, the (\ref{strong}) channel dominates
over the (\ref{pn}) one when the proton acceleration is much larger than the
pion mass~\cite{TK}. The same sort of approximations can be used to investigate
Eq.~(\ref{strong}) but unfortunately the results obtained are only good when the
proton acceleration is large enough for the pion to be assumed massless.  
The development of more powerful approximations able to investigate process 
(\ref{strong}) 
in a larger acceleration range  (including $a \approx m_{\pi}$) would be 
welcome. 
A comprehensive analysis of the consequences of the proton decay to 
astrophysics is being  considered.

\begin{flushleft}
{\bf{\large Acknowledgments}}
\end{flushleft}

G.M. is thankful to Conselho Nacional de Desenvolvimento Cient\'\i fico e 
Tecnol\'ogico for partial support while D.V. is grateful to the US 
National Science Foundation for full support under Grant  No PHY-0071044.

%%%%%%%%%%%%%%%%%%%%%%%%%%%%%%%%%%%%%%%%%%%%%%%%%%%%%%%%%%%%%%%%%%%%%%%%
%				BIBLIOGRAPHY
%%%%%%%%%%%%%%%%%%%%%%%%%%%%%%%%%%%%%%%%%%%%%%%%%%%%%%%%%%%%%%%%%%%%%%%%

\end{document}